# Reactions of $C_2(a^3\Pi_u)$ with selected saturated alkanes: A temperature dependence study

**Renzhi Hu, Qun Zhang,[a] and Yang Chen[a]**

*Hefei National Laboratory for Physical Sciences at the Microscale and Department of Chemical Physics, University of Science and Technology of China, Hefei, Anhui 230026, People's Republic of China*

[a]Authors to whom correspondence should be addressed. Electronic addresses: qunzh@ustc.edu.cn and yangchen@ustc.edu.cn

## ABSTRACT

We present a temperature dependence study on the gas phase reactions of the $C_2(a^3\Pi_u)$ radical with a selected series of saturated alkanes ($C_2H_6$, $C_3H_8$, $n$-$C_4H_{10}$, $i$-$C_4H_{10}$, and $n$-$C_6H_{14}$) by means of pulsed laser photolysis/laser-induced fluorescence technique. The bimolecular rate constants for these reactions were obtained between 298 and 673 K. A pronounced negative temperature effect was observed for $n$-$C_4H_{10}$, $i$-$C_4H_{10}$, and $n$-$C_6H_{14}$ and interpreted in terms of steric hindrance of the more reactive secondary or tertiary C-H bonds by less reactive $CH_3$ groups. Detailed analysis of our experimental results reveals quantitatively the temperature dependence of reactivities for the primary, secondary, and tertiary C-H bonds in these saturated alkanes and further lends support to a mechanism of hydrogen abstraction.

## I. INTRODUCTION

In light of its importance in a wide variety of natural and man-made environments such as interstellar clouds, planetary atmospheres, plasmas, combustion flames, and various chemical systems, the gas phase kinetics of reactions of the dicarbon molecule, $C_2$, with other co-reagent species has received increased attention during the past decade.[1-12] While growing interest in the temperature dependence for the reactions involving $C_2$ (in its ground singlet $X^1\Sigma_g^+$ state and/or its first excited triplet $a^3\Pi_u$ state) has since 2006 turned to the low temperature limit,[1-4] it is surprising that the relevant data available in the high temperature regime remain incomplete. Among the large amount of earlier kinetic studies performed at ambient or higher temperatures on the reactions of $C_2$ (in its $X^1\Sigma_g^+$ or $a^3\Pi_u$ states) with other co-reagent species (mainly including small inorganic species and some hydrocarbons),[5-22] not many have addressed the temperature dependence issue.[9-12,20-22] Ristanovic *et al.*[9] studied the $C_2(a^3\Pi_u)$ + $NO(X^2\Pi)$ → $CN(A^2\Pi, B^2\Sigma)$ + $CO(X^1\Sigma)$ reaction in the temperature range 292 – 968 K; by comparing their results to the $C_2$ + NO measurements by Kruse and Roth[12] for the temperature range 3150 – 3950 K, they confirmed that a mechanism change occurs between the two temperature ranges. Fontijn *et al.*[10] studied the $C_2(X^1\Sigma_g^+, a^3\Pi_u)$ + $O_2(X^3\Sigma)$ reaction in the temperature range 298 – 711 K. Pitts *et al.*[22] investigated the same reaction as well as the $C_2(X^1\Sigma_g^+)$ + $H_2$ and $CH_4$ in the temperature range 300 – 600 K. Becker *et al.*[11] studied the $C_2(a^3\Pi_u)$ + $C_2H_4$ and $N_2O$ reactions in the temperature ranges 298 – 1000 K and 300 – 700 K, respectively. Pasternack *et al.*[21] reported the disappearance rates of $C_2(a^3\Pi_u)$ in the temperature range 300 – 600 K for reactions with $H_2$, $D_2$, $C_2H_6$, $C_3H_8$, $n$-$C_4H_{10}$, and $C_2H_4$ and intersystem crossing due to collisions with Xe.

The temperature-dependent reaction kinetics of $C_2(a^3\Pi_u)$ with saturated alkanes of different lengths, to the best of our knowledge, have hitherto only been reported by Pasternack *et al.*,[20,21] who measured the rate constants $k$ (in the units cm$^3$ molecule$^{-1}$ s$^{-1}$) of the $C_2(a^3\Pi_u)$ reactions with CH$_4$ in 337 − 605 K giving $k(CH_4) = (1.65 \pm 0.20) \times 10^{-11}$ exp(-(2803 ± 55)/$T$)[20] and with C$_2$H$_6$ in 300 − 524 K, C$_3$H$_8$ in 299 − 441 K, and $n$-C$_4$H$_{10}$ in 298 − 480 K giving $k(C_2H_6) = (2.42 \pm 0.10) \times 10^{-11}$ exp(-(919 ± 15)/$T$), $k(C_3H_8) = (1.84 \pm 0.17) \times 10^{-11}$ exp(-(97 ± 36)/$T$), and $k(n$-C$_4$H$_{10}) = (4.9 \pm 0.5) \times 10^{-11}$ exp(-(71 ± 41)/$T$), respectively.[21] However, since at only three temperatures they measured $k$ constants for C$_3$H$_8$ and $n$-C$_4$H$_{10}$, their corresponding fits in Arrhenius form seem not reliable, as can be seen from the large uncertainties of the activation energy. By correlating their temperature dependence data using the bond energy-bond order (BEBO) calculations,[23] linear free energy correlations,[24] and Evans-Polanyi plots,[25] they drew a conclusion that the reactions of $C_2(a^3\Pi_u)$ + CH$_4$, C$_2$H$_6$, C$_3$H$_8$, and $n$-C$_4$H$_{10}$ yielding C$_2$H most likely proceed *via* a hydrogen abstraction mechanism[21] instead of an insertion mechanism.[19] In our previous work,[8] we investigated the reactions of $C_2(a^3\Pi_u)$ with (C$_1$ − C$_8$) alkanes at 300 K and observed a correlation between the reaction rate and the C-chain length of the alkanes, which was found to be in favor of the H-atom abstraction mechanism proposed by Pasternack *et al.*[21]

We present here a more systematic and detailed temperature dependence study on the reactions of the $C_2(a^3\Pi_u)$ radical with a selected series of saturated alkane molecules including C$_2$H$_6$, C$_3$H$_8$, $n$-C$_4$H$_{10}$, $i$-C$_4$H$_{10}$, and $n$-C$_6$H$_{14}$ over the temperature range 298 − 673 K, from which we found a pronounced negative temperature dependence of the rate constants for the reactions of $C_2(a^3\Pi_u)$ with $n$-$C_4H_{10}$, $i$-$C_4H_{10}$, and $n$-$C_6H_{14}$ below ~373 K. In addition, we obtained the estimated rate constants pertinent to the attack of $C_2(a^3\Pi_u)$ on each secondary C-H bond in $C_3H_8$, $n$-$C_4H_{10}$, and $n$-$C_6H_{14}$, as well as those pertinent to the attack of $C_2(a^3\Pi_u)$ on the tertiary C-H bond in $i$-$C_4H_{10}$, based on which we further confirmed that the reactions of $C_2(a^3\Pi_u)$ with the saturated alkane molecules proceed *via* a hydrogen abstraction mechanism.

## II. EXPERIMENT

The experimental apparatus is similar to that used in our previous studies on the $C_2(a^3\Pi_u)$ radical.[5-8] The pulsed laser photolysis/laser induced fluorescence experiments were performed in a stainless steel flow reactor. The $C_2(a^3\Pi_u)$ radicals were produced by photolysis of $C_2Cl_4$ using the fourth harmonic output (266 nm, typical energy ~5.5 mJ/pulse) of a Nd:YAG laser (INDI, Spectra Physics) operating at a repetition rate of 10 Hz. A lens with a 500-mm focal length was used to focus the photolysis laser beam into the center of the reaction cell. The probe laser used was a Nd:YAG laser (GCR-170, Spectra Physics) pumped dye laser (PrecisionScan, Sirah) which outputs a ~1.5 mJ pulse energy and operates at a repetition rate of 10 Hz. The dye laser beam was set collinear to the photolysis laser beam in a counter-propagating way. To effectively suppress the scattered light, we confined the diameter of the dye laser beam to ~1 mm.

The $C_2(a^3\Pi_u)$ concentrations were probed by exciting the $d^3\Pi_g \leftarrow a^3\Pi_u$ 0-0 transition at 516.5 nm[26] and detecting the fluorescence on the 0-1 vibronic transition band which was isolated with an interference filter at 563.5 nm. The LIF signal was collected by a photomultiplier tube (R928, Hamamatsu) whose output was recorded by a digital oscilloscope (TDS380, Tektronix) and then averaged over 256 laser pulses with a computerized data acquisition system. A digital delay generator (DG535, Stanford Research) was used to vary the time delay between the photolysis and probe laser pulses.

Temperatures of the resistively heated flow reactor were controlled over the range 298 − 673 K by a temperature regulator (AI-708P, Xiamen) and measured by a K-type thermocouple probe mounted a few millimeters away from the center of the reactor. The controlled temperatures were found to be within a ±0.5 K precision over the dimensions of the probed volume and the duration of our experiments.

In a typical experiment, three gas flows including premixed $C_2Cl_4$/Ar, alkane/Ar, and pure argon gas slowly passed through the reaction cell and were measured independently by three calibrated mass flow controllers (D07-7A/2M, Beijing). The concentrations of alkane reactants can be calculated by

$$[\text{alkane}] = 9.66 \times 10^{18} \frac{p}{T} \frac{f_{\text{alkane/Ar}} x_{\text{alkane/Ar}}}{f_{\text{total}}}, \tag{1}$$

where [alkane] is the concentration of alkane reactant (in molecules cm$^{-3}$), $p$ and $T$ are the pressure (Torr) and temperature (K) of the system, $f_{\text{alkane/Ar}}$ and $f_{\text{total}}$ are the flow rates of alkane/Ar and all gases, respectively, and $x_{\text{alkane/Ar}}$ is the pressure ratio of alkane over Ar. All experiments were performed over the temperature range 298 − 673K and at a total pressure of ~9.5 Torr to maintain a steady flow condition.

The $C_2(a^3\Pi_u)$ radicals generated from the 266 nm laser photolysis of the precursor $C_2Cl_4$ molecules are rotationally hot.[5-8] The Ar buffer gas was included in the reaction mixture in order to relax the nascent quantum state distributions of both the reactants and the $C_2(a^3\Pi_u)$ radicals as well as to slow down the diffusion of molecules out of the detection region. The LIF decay of $C_2(a^3\Pi_u)$ was measured after $C_2(a^3\Pi_u)$ is rotationally cooled to room temperature. Under the above experimental conditions, since the $C_2(a^3\Pi_u)$ radicals were found to be rotationally cooled to the room temperature within a ~8 μs time delay between the photolysis and probe laser pulses,[5-8] we collected the kinetic rate data thereafter.

The materials used in the experiments were as follows: $C_2Cl_4$ (>97%, Shanghai), ethane (99.99%, Nanjing), propane (99.9%, Nanjing), *n*-butane (99.9%, Nanjing), *i*-butane (99.99%, Nanjing), and *n*-hexane (>99.9%, Tianjin). All these reagents were degassed by three freeze-pump-thaw cycles in liquid nitrogen. Ar (99.999%, Nanjing) was used without further purification.

## III. RESULTS AND DISCUSSION

Similar to our previous work,[5-8] the validity of pseudo-first-order kinetics was ensured throughout all the measurements we present here by keeping the partial pressure of the $C_2Cl_4$ precursor always much lower than that of the alkane reactants. Under such a condition, the LIF decay of $C_2(a^3\Pi_u)$ at long times (>8 μs) follows

$$I = I_0 \exp(-k't), \qquad (2)$$

where $k'$ is the pseudo-first-order rate constant for the total loss of the $C_2(a^3\Pi_u)$ LIF signal due to all processes including reactions and diffusion out of the probe region, and $t$ is the time delay between the photolysis and probe lasers. Figure 1 shows a typical LIF decay trace for the case of $C_3H_8$ at 473 K. At various reactant concentrations we obtained the corresponding $k'$ values through exponential fits using Eq. (2).

As a result of the pseudo-first-order kinetics, $k'$ at a certain concentration of alkane under investigation is given by

$$k' = k[\text{alkane}] + k_n, \qquad (3)$$

where $k$ is the bimolecular rate constant for the reaction of $C_2(a^3\Pi_u)$ with alkane, and $k_n$ is the loss rate constant for $C_2(a^3\Pi_u)$ due to the reactions and diffusion in the absence of alkane. Linear least-squares fit of $k'$ versus [alkane] yields therefore the bimolecular rate constant $k$ for the reaction of $C_2(a^3\Pi_u)$ with alkane. Figure 2, as an example, shows plots of $k'$ as a function of the concentrations of $C_2H_6$, $C_3H_8$, and $n$-$C_4H_{10}$ at 473K, 473K, and 523K, respectively, all of which exhibit excellent linear relationship.

Table I lists the obtained bimolecular rate constants (in $10^{-11}$ cm$^3$ molecule$^{-1}$ s$^{-1}$, with $\pm 2\sigma$ errors) for the reactions of $C_2(a^3\Pi_u)$ with the selected alkanes ($C_2H_6$, $C_3H_8$, $n$-$C_4H_{10}$, $i$-$C_4H_{10}$, and $n$-$C_6H_{14}$) over the temperature range 298 − 673K.

The temperature dependence of $k(C_2H_6)$ is displayed in Figure 3(a) and in Arrhenius form in Figure 3(b). The solid curve in Fig. 3(a) is a least-squares fit of the measured rate constants using the modified three-parameter Arrhenius equation

$$k = AT^B \exp(-C/T), \qquad (4)$$

which leads to

$$k(C_2H_6) = (1.01 \pm 2.28) \times 10^{-13}\, T^{1.13 \pm 0.31} \exp(-(482.21 \pm 155.88)/T). \qquad (5)$$

It is obvious that the parameter $A$ in Eq. (5) gives an incredible uncertainty that exceeds 200%, which implies that employing here a three-parameter expression for the temperature dependence of $k(C_2H_6)$ is inappropriate. Nevertheless, a fit in normal Arrhenius form turns out a reasonable result:

$$k(C_2H_6) = (3.04 \pm 0.07) \times 10^{-11} \exp(-(970.82 \pm 9.19)/T), \quad (6)$$

as plotted as the solid line on a semi-logarithm scale in Fig. 3(b). We also included in Figs. 3(a) and (b), for comparison, the measured data previously reported,[1,14,21] from which we can see that (1) our measurements are in reasonably good agreement with the data given in Ref. 21 (shown as the five filled symbols in Figs. 3(a) and (b)) and (2) our room-temperature measurement bears the smallest uncertainty compared to the other three,[1,14,21] as shown in the insets of Figs. 3(a) and (b).

In contrast to the case of $C_2H_6$, we found that for the remaining reactions (*i.e.*, $C_2(a^3\Pi_u)$ + $C_3H_8$, $n$-$C_4H_{10}$, $i$-$C_4H_{10}$, and $n$-$C_6H_{14}$), the temperature dependence of $k$ all shows strongly curved non-Arrhenius behavior, as illustrated respectively in Figures 4 − 7. Since the fit in normal Arrhenius form (ln$k$ *versus* 1/$T$) turned out not to yield a reasonable linear relationship, we used here the conventionally adopted three-parameter Arrhenius form (*i.e.*, Eq. (4)) to describe the temperature dependence of $k$ for the four reactions. The results are as follows:

$$k(C_3H_8) = (7.90 \pm 2.62) \times 10^{-19} T^{2.44 \pm 0.04} \exp((811.46 \pm 21.24)/T), \quad (7)$$

$$k(n\text{-}C_4H_{10}) = (1.96 \pm 0.24) \times 10^{-16} T^{1.73 \pm 0.02} \exp((630.51 \pm 7.53)/T), \quad (8)$$

$$k(i\text{-}C_4H_{10}) = (9.30 \pm 1.36) \times 10^{-15} T^{1.30 \pm 0.02} \exp((483.07 \pm 9.14)/T), \quad (9)$$

$$k(n\text{-}C_6H_{14}) = (1.64 \pm 0.57) \times 10^{-19} T^{2.70 \pm 0.05} \exp((1021.71 \pm 22.29)/T), \quad (10)$$

and correspondingly plotted as the solid curves in Figs. 4 − 7. It is apparent that the curvatures of the four curves are rather different, the reason of which will be explained later. Another noteworthy point is the appearance of "negative activation energy" in Eqs. (7) − (10), which is commonly accounted for the existence of a strong long-range attractive interaction between the reactants as well as the absence of an appreciable reaction barrier.[27]

The available data measured previously[1,8,21] pertinent to the four reaction systems are also included in Figs. 4 − 7. It can be readily seen that our measurements given in Fig. 4, similar to the case of $C_2H_6$, are in reasonably good agreement with the data given in Ref. 21 (shown as the three filled symbols in Fig. 4), while in Fig. 5 the three measured data reported by the authors of Ref. 21 are all ~20% larger than our data measured at almost identical temperatures. Such a discrepancy, which is usually regarded as being tolerable in kinetic rate measurements,[28] may result from different experimental conditions; nevertheless, the trend of $k$ varying with $T$ in both measurements appears quite similar. In Figs. 6 and 7, apart from the two room-temperature data adapted from Ref. 8, the results for $i$-$C_4H_{10}$ (Fig. 6) and $n$-$C_6H_{14}$ (Fig. 7) fitted respectively to Eqs. (9) and (10) are all reported for the first time. Again, as in the case of $C_2H_6$, the uncertainty of the room-temperature data we measured for each of the four reactions is the smallest, hence more reliable, compared to the available data previously reported,[1,8,21] as shown in Figs. 4 − 7.

From Figs. 5 − 7 one can easily detect a pronounced negative temperature dependence of the bimolecular rate constants for $n$-$C_4H_{10}$, $i$-$C_4H_{10}$, and $n$-$C_6H_{14}$ below ~373 K (*i.e.*, $1000/T$ > ~2.7 in Figs. 5 − 7). Such a negative temperature effect, as discussed in a previous study on the temperature dependence of CN reactions with selected alkanes,[27] may be mainly attributed to steric hindrance of the more reactive secondary (for $n$-$C_4H_{10}$ and $n$-$C_6H_{14}$) or tertiary (for $i$-$C_4H_{10}$) C-H bonds by less reactive $CH_3$ groups. Dynamically, rotations of the $n$-$C_4H_{10}$, $i$-$C_4H_{10}$, and $n$-$C_6H_{14}$ molecules and their $CH_3$ groups become slower with temperature decreasing from 373 to 298 K, which allows $C_2(a^3\Pi_u)$ to reach the secondary or tertiary H atoms more easily and thereby make the rate constants increase. This can be valid only when the energy barrier for the $C_2(a^3\Pi_u)$ attack on the secondary and tertiary C-H bonds is negligibly small and that for the attack on the primary C-H bonds is not negligible.[27]

Compared to the cases of $n$-$C_4H_{10}$, $i$-$C_4H_{10}$, and $n$-$C_6H_{14}$ (Figs. 5 − 7), $C_2(a^3\Pi_u)$ + $C_3H_8$ shows a much weaker negative temperature effect to the low temperature limit, as can be discerned as a slight upward curvature below ~323 K in Fig. 4. We expect that it may become more appreciable if extended to lower temperatures; this expectation has, as a matter of fact, been verified by a very recent low-temperature work whose data measured in 300 – 83 K (together with our data presented in Fig. 4) are plotted in the inset of Fig. 4.

To gain more insights into the temperature dependence, we further examined the reactivities for the three types of C-H bonds (*i.e.*, the primary, secondary, and tertiary C-H bonds).

As in Refs. 27 and 8, the measured rate constant for $C_2(a^3\Pi_u)$ + alkane can be reasonably approximated as the sum of the rate constants for the $C_2(a^3\Pi_u)$ reaction with the three types of C-H bonds in each alkane, *e.g.*, that for $C_2(a^3\Pi_u)$ + $C_2H_6$ represents six primary C-H bond reaction rate constants, while that for $C_2(a^3\Pi_u)$ + $n$-$C_4H_{10}$ represents six primary and four secondary C-H bond reaction rate constants. This can be summarized as

$$k = N_p k_p + N_s k_s + N_t k_t, \qquad (11)$$

where $k$ is the overall rate constant for the alkane under investigation, $k_p$, $k_s$, and $k_t$ are the corresponding rate constants for the reaction of $C_2(a^3\Pi_u)$ with the primary, secondary, and tertiary C-H bonds, respectively, while $N_p$, $N_s$, and $N_t$ are the corresponding C-H bond numbers. Considering that all the primary C-H bonds of saturated alkanes locate at the ends of C-chain and hence their reactivities with $C_2(a^3\Pi_u)$ in differing saturated alkanes can be thought of as to be similar, we can assume that $k_p$ in the five alkane molecules under investigation is essentially identical at a certain temperature.[27-30] Thus we have

$$k_p \equiv k_p(C_2H_6) = k(C_2H_6)/6, \quad (12)$$

$$k_s(C_3H_8) = (k(C_3H_8) - 6k_p)/2, \quad (13)$$

$$k_s(n\text{-}C_4H_{10}) = (k(n\text{-}C_4H_{10}) - 6k_p)/4, \quad (14)$$

$$k_s(n\text{-}C_6H_{14}) = (k(n\text{-}C_6H_{14}) - 6k_p)/8, \quad (15)$$

$$k_t(i\text{-}C_4H_{10}) = k(i\text{-}C_4H_{10}) - 9k_p. \quad (16)$$

The $k_p$ and $k_t(i\text{-}C_4H_{10})$ data as a function of temperature are plotted in Figures 8(a) and (b), respectively, while Figure 8(c) plots the $k_s(C_3H_8)$, $k_s(n\text{-}C_4H_{10})$, and $k_s(n\text{-}C_6H_{14})$ data labeled (i), (ii), and (iii), respectively. The temperature dependences of $k_t(i\text{-}C_4H_{10})$, $k_s(C_3H_8)$, $k_s(n\text{-}C_4H_{10})$, and $k_s(n\text{-}C_6H_{14})$ fitted by the three-parameter Arrhenius form are as follows:

$$k_t(i\text{-}C_4H_{10}) = (5.05 \pm 2.42) \times 10^{-18} T^{2.12 \pm 0.07} \exp((966.07 \pm 29.67)/T), \quad (17)$$

$$k_s(C_3H_8) = (5.38 \pm 3.03) \times 10^{-18} T^{1.98 \pm 0.07} \exp((794.50 \pm 35.19)/T), \quad (18)$$

$$k_s(n\text{-}C_4H_{10}) = (4.66 \pm 0.98) \times 10^{-16} T^{1.37 \pm 0.03} \exp((564.16 \pm 13.11)/T), \quad (19)$$

$$k_s(n\text{-}C_6H_{14}) = (3.32 \pm 0.58) \times 10^{-15} T^{1.14 \pm 0.02} \exp((447.97 \pm 10.81)/T), \quad (20)$$

and correspondingly plotted as the solid curves in Figs. 8(b) and (c).

From Fig. 8, we can arrive at four main points as follows: (1) $k_t(i\text{-}C_4H_{10}) > k_s(C_3H_8, n\text{-}C_4H_{10},$ and $n\text{-}C_6H_{14}) \sim 10k_p$, which actually reveals the reactivities for the three types of C-H bonds accordingly in these saturated alkanes and hence further supports the mechanism of hydrogen abstraction in the reactions of $C_2(a^3\Pi_u)$ with the saturated alkanes;[8,21] (2) $k_s(C_3H_8) < k_s(n\text{-}C_4H_{10}) < k_s(n\text{-}C_6H_{14})$, which may result from the fact that the secondary C-H bonds for the three species reside in distinct chemical environments; (3) The $k_p$ curve (in Fig. 8(a)) decays monotonically with decreasing temperature while the $k_t$ curve (in Fig. 8(b)) and the three $k_s$ curves (in Fig. 8(c)) all exhibit non-monotonic behavior, which implies that only secondary and tertiary C-H bonds contribute significantly to the observed negative temperature effect, as discussed above; (4) The curvatures for the $k_s$ curves are almost the same but clearly smaller than that for the $k_t$ curve, which simply tells us that the tertiary C-H bond > the secondary C-H bond in terms of contributions to the negative temperature effect. It is points (3) and (4) (combined with an explicit consideration of the secondary C-H bond numbers in $C_3H_8$, $n$-$C_4H_{10}$, and $n$-$C_6H_{14}$) that lead to the aforementioned curvature difference for the curves shown in Figs. 4 – 7.

## V. CONCLUSIONS

The bimolecular rate constants for the reactions of $C_2(a^3\Pi_u)$ with $C_2H_6$, $C_3H_8$, $n$-$C_4H_{10}$, $i$-$C_4H_{10}$, and $n$-$C_6H_{14}$ have been measured over the temperature range 298 − 673 K by means of laser photolysis/laser-induced fluorescence technique. These rate constants, expressed in the units $cm^3$ $molecule^{-1}$ $s^{-1}$, can be effectively represented by $k(C_2H_6) = (3.04 \pm 0.07) \times 10^{-11}$ $\exp(-(970.82 \pm 9.19)/T)$, $k(C_3H_8) = (7.90 \pm 2.62) \times 10^{-19} T^{2.44 \pm 0.04} \exp((811.46 \pm 21.24)/T)$, $k(n$-$C_4H_{10}) = (1.96 \pm 0.24) \times 10^{-16} T^{1.73 \pm 0.02} \exp((630.51 \pm 7.53)/T)$, $k(i$-$C_4H_{10}) = (9.30 \pm 1.36) \times 10^{-15} T^{1.30 \pm 0.02} \exp((483.07 \pm 9.14)/T)$, and $k(n$-$C_6H_{14}) = (1.64 \pm 0.57) \times 10^{-19} T^{2.70 \pm 0.05} \exp((1021.71 \pm 22.29)/T)$. Pronounced negative temperature dependences of the rate constants for the reactions of $C_2(a^3\Pi_u)$ with $n$-$C_4H_{10}$, $i$-$C_4H_{10}$, and $n$-$C_6H_{14}$ have been detected below ~373 K, which is mainly attributed to steric hindrance of the more reactive secondary or tertiary C-H bonds by less reactive $CH_3$ groups.

Based on the rate constants of $C_2H_6$ which arise only from the primary C-H bond contributions, we have estimated the rate constants (in the units $cm^3$ $molecule^{-1}$ $s^{-1}$) pertinent to the attack of $C_2(a^3\Pi_u)$ on each secondary C-H bond in $C_3H_8$, $n$-$C_4H_{10}$, and $n$-$C_6H_{14}$ to be, respectively, $k_s(C_3H_8) = (5.38 \pm 3.03) \times 10^{-18} T^{1.98 \pm 0.07} \exp((794.50 \pm 35.19)/T)$, $k_s(n$-$C_4H_{10}) = (4.66 \pm 0.98) \times 10^{-16} T^{1.37 \pm 0.03} \exp((564.16 \pm 13.11)/T)$, and $k_s(n$-$C_6H_{14}) = (3.32 \pm 0.58) \times 10^{-15} T^{1.14 \pm 0.02} \exp((447.97 \pm 10.81)/T)$, as well as those pertinent to the attack of $C_2(a^3\Pi_u)$ on the tertiary C-H bond in $i$-$C_4H_{10}$ to be $k_t(i$-$C_4H_{10}) = (5.05 \pm 2.42) \times 10^{-18} T^{2.12 \pm 0.07} \exp((966.07 \pm 29.67)/T)$. We also found that only secondary and tertiary C-H bonds contribute significantly to the observed negative temperature effect. In addition, our experimental results showing $k_t(i$-$C_4H_{10}) > k_s(C_3H_8$, $n$-$C_4H_{10}$, and $n$-$C_6H_{14}) \sim 10 k_p$ support the mechanism of hydrogen abstraction in the reactions of $C_2(a^3\Pi_u)$ with the saturated alkanes.

## ACKNOWLEDGEMENTS

We gratefully acknowledge support from the National Natural Science Foundation of China (Grant Nos. 20673107 and 20873133), the Ministry of Science and Technology of China (Grant Nos. 2007CB815203 and 2010CB923302), the Chinese Academy of Sciences (KJCX2-YW-N24), and the Scientific Research Foundation for the Returned Overseas Chinese Scholars, Ministry of Education of China.

# Figure Captions

FIG. 1.   Typical LIF decay trace for $C_2(a^3\Pi_u)$ in the presence of $C_2Cl_4$, $C_3H_8$, and excess Ar buffer gas with a total pressure of ~9.5 Torr at 473 K.   $[C_2Cl_4] = 9.2 \times 10^{13}$ molecules $cm^{-3}$ and $[C_3H_8] = 7.5 \times 10^{14}$ molecules $cm^{-3}$.   The solid curve is an exponential fit of the experimental data, which yields a pseudo-first-order rate constant $k'$.

FIG. 2.   Plots of pseudo-first-order rate constants versus the concentrations of $C_2H_6$, $C_3H_8$, and $n$-$C_4H_{10}$ at 473K, 473K, and 523K, respectively, in the presence of excess Ar buffer gas with a total pressure of ~9.5 Torr.   The solid lines are linear least-squares fits of the experimental data.

FIG. 3.   (Color online) Bimolecular rate constants for the reaction of $C_2(a^3\Pi_u) + C_2H_6$ as a function of temperature over the range 298 – 673 K.   (a) Plot in three-parameter form; (b) Plot in Arrhenius form (displayed on a semi-logarithm scale).   The error bars represent ±2σ estimates of the total experimental error.   The solid curves are least-squares fits of our experimental data.   The insets are expanded view of the data in the oval.

FIG. 4.   (Color online) Bimolecular rate constants for the reaction of $C_2(a^3\Pi_u) + C_3H_8$ as a function of temperature over the range 298 – 673 K.   The error bars represent ±2σ estimates of the total experimental error.   The solid curve is a three-parameter least-squares fit of our experimental data.   The inset shows the data measured in 300 – 83 K (adapted from Ref. 1)

together with our data measured in 298 – 673 K.

FIG. 5 (Color online) Bimolecular rate constants for the reaction of $C_2(a^3\Pi_u)$ + $n$-$C_4H_{10}$ as a function of temperature over the range 298 – 673 K. The error bars represent ±2σ estimates of the total experimental error. The solid curve is a three-parameter least-squares fit of our experimental data.

FIG. 6. (Color online) Bimolecular rate constants for the reaction of $C_2(a^3\Pi_u)$ + $i$-$C_4H_{10}$ as a function of temperature over the range 298 – 673 K. The error bars represent ±2σ estimates of the total experimental error. The solid curve is a three-parameter least-squares fit of our experimental data.

FIG. 7. (Color online) Bimolecular rate constants for the reaction of $C_2(a^3\Pi_u)$ + $n$-$C_6H_{14}$ as a function of temperature over the range 298 – 673 K. The error bars represent ±2σ estimates of the total experimental error. The solid curve is a three-parameter least-squares fit of our experimental data.

FIG. 8. Bimolecular rate constants for the reactions of $C_2(a^3\Pi_u)$ with (a) the primary C-H bond in $C_2H_6$, (b) the tertiary C-H bond in $i$-$C_4H_{10}$, and (c) the secondary C-H bonds in (i) $C_3H_8$, (ii) $n$-$C_4H_{10}$, and (iii) $n$-$C_6H_{14}$, as a function of temperature over the range 298 – 673 K. The error bars represent ±2σ estimates of the total experimental error. The solid curves are three-parameter least-squares fits of the experimental data.

**Fig. 1 (Hu *et al.*)**

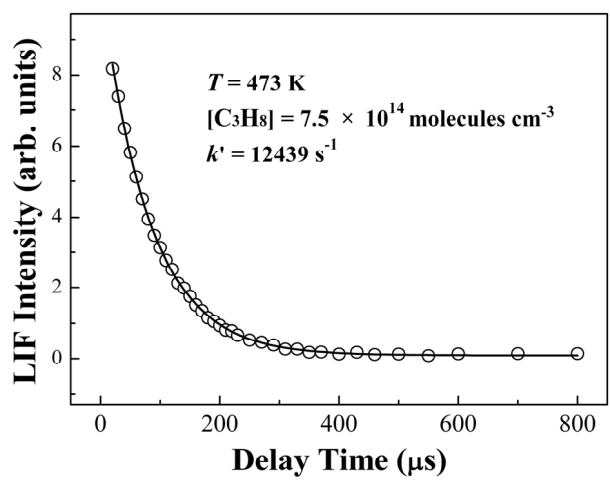

**Fig. 2** (Hu *et al.*)

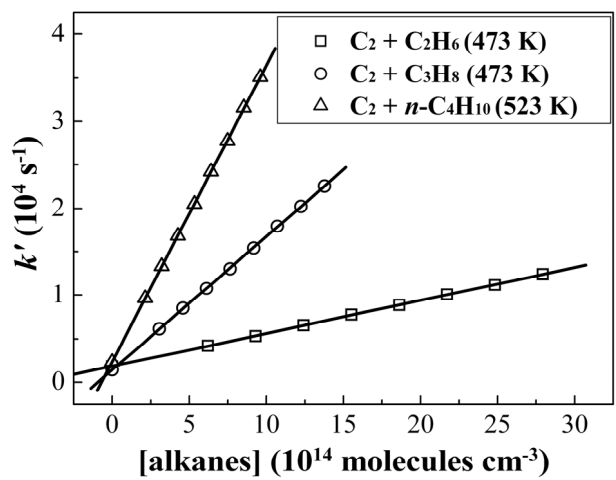

**Fig. 3 (Hu *et al.*)**

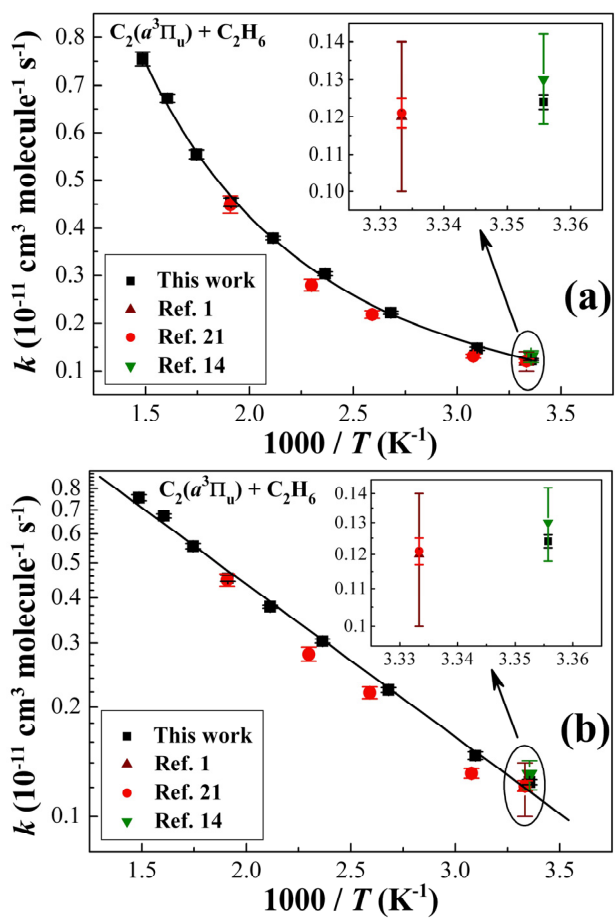

**Fig. 4   (Hu *et al.*)**

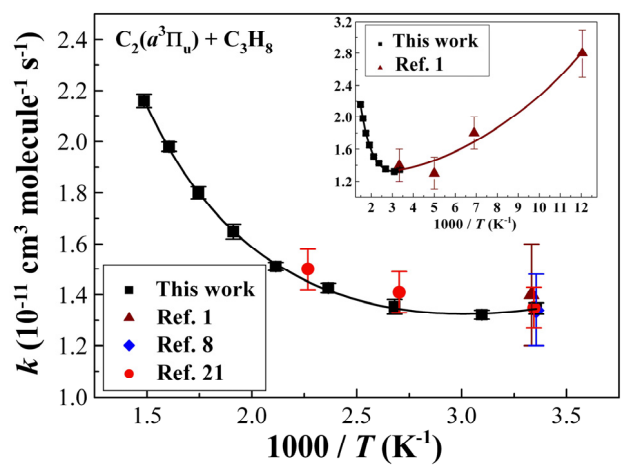

**Fig. 5  (Hu *et al.*)**

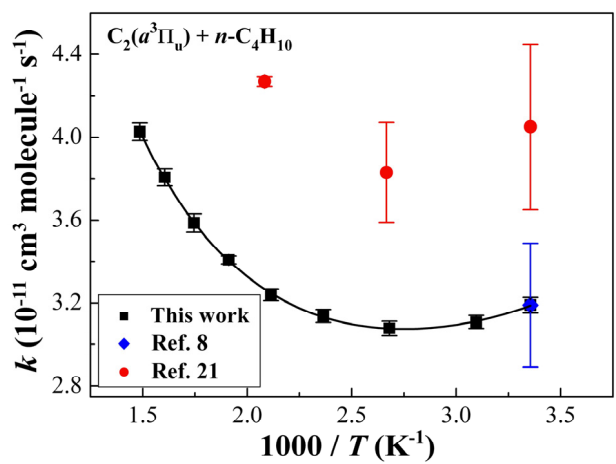

**Fig. 6 (Hu *et al.*)**

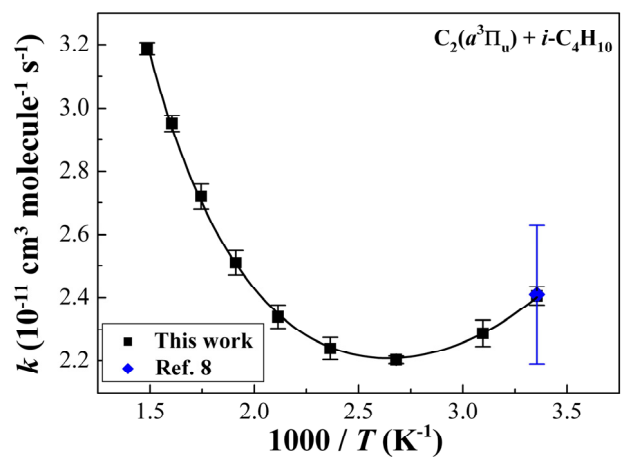

**Fig. 7   (Hu *et al.*)**

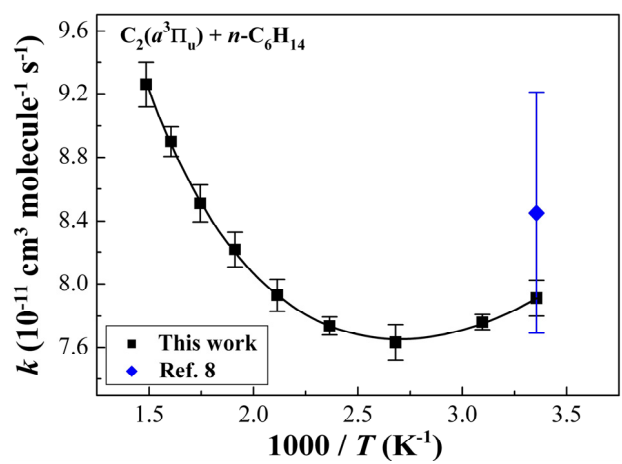

**Fig. 8** (Hu *et al.*)

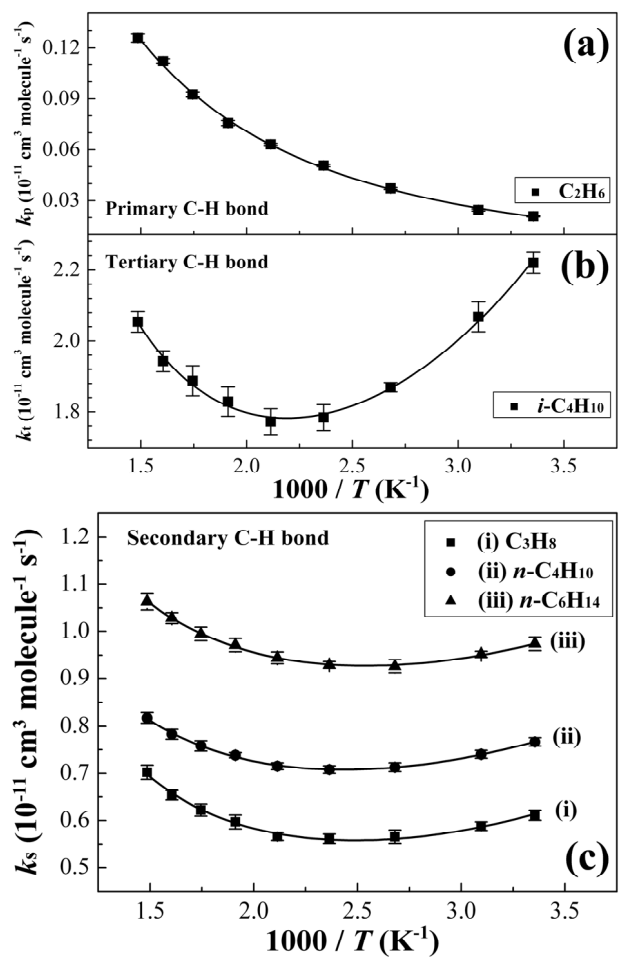

TABLE I  Bimolecular rate constants $k$ (in $10^{-11}$ cm$^3$ molecule$^{-1}$ s$^{-1}$) for reactions of $C_2(a^3\Pi_u)$ with $C_2H_6$, $C_3H_8$, $n$-$C_4H_{10}$, $i$-$C_4H_{10}$, and $n$-$C_6H_{14}$ over the temperature range 298 – 673 K.  The error bars represent ±2σ estimates of the total experimental error.

| $T$ (K) | $C_2H_6$ | $C_3H_8$ | $n$-$C_4H_{10}$ | $i$-$C_4H_{10}$ | $n$-$C_6H_{14}$ |
|---|---|---|---|---|---|
| 298 | 0.124 ± 0.002 | 1.35 ± 0.02 | 3.19 ± 0.04 | 2.41 ± 0.03 | 7.91 ± 0.11 |
| 323 | 0.147 ± 0.004 | 1.32 ± 0.02 | 3.11 ± 0.03 | 2.29 ± 0.04 | 7.76 ± 0.05 |
| 373 | 0.223 ± 0.003 | 1.35 ± 0.03 | 3.08 ± 0.04 | 2.20 ± 0.01 | 7.63 ± 0.11 |
| 423 | 0.304 ± 0.004 | 1.43 ± 0.02 | 3.14 ± 0.03 | 2.24 ± 0.04 | 7.74 ± 0.06 |
| 473 | 0.379 ± 0.004 | 1.51 ± 0.02 | 3.24 ± 0.03 | 2.34 ± 0.04 | 7.93 ± 0.10 |
| 523 | 0.454 ± 0.009 | 1.65 ± 0.03 | 3.41 ± 0.02 | 2.51 ± 0.04 | 8.22 ± 0.11 |
| 573 | 0.555 ± 0.009 | 1.80 ± 0.02 | 3.59 ± 0.04 | 2.72 ± 0.04 | 8.51 ± 0.11 |
| 623 | 0.673 ± 0.002 | 1.98 ± 0.02 | 3.80 ± 0.04 | 2.95 ± 0.03 | 8.90 ± 0.09 |
| 673 | 0.755 ± 0.014 | 2.16 ± 0.03 | 4.03 ± 0.04 | 3.19 ± 0.02 | 9.26 ± 0.14 |